\documentclass[twocolumn,prl,aps,floatfix]{revtex4}
\usepackage{psfig}

 \newcommand\la{\langle}
 \newcommand\ra{\rangle}
 \newcommand\beq{\begin{equation}}
 
 \newcommand\eeq{\end{equation}}                                               
 \newcommand\beqn{\begin{eqnarray}}
 \newcommand\eeqn{\end{eqnarray}}

\def\mb{\,\mbox{mb}}
\def\fm{\,\mbox{fm}}
\def\GeV{\,\mbox{GeV}}
\def\TeV{\,\mbox{TeV}}

\def\lsim{\mathrel{\rlap{\lower4pt\hbox{\hskip1pt$\sim$}}
    \raise1pt\hbox{$<$}}}         
\def\gsim{\mathrel{\rlap{\lower4pt\hbox{\hskip1pt$\sim$}}
    \raise1pt\hbox{$>$}}}         

\begin{document} 
 
\title{Cronin Effect in Hadron Production off Nuclei}

\author{ 
B.Z.~Kopeliovich$^{1,2,3}$, J.~Nemchik$^{4}$, 
A.~Sch\"afer$^{2}$ and A.V.~Tarasov$^{1,2,3}$}
\affiliation{ $^{1}$ Max-Planck Institut f\"ur Kernphysik, Postfach 
103980, 69029 Heidelberg, Germany\\
 $^{2}$ Institut f\"ur Theoretische Physik der Universit\"at, 93040
Regensburg, Germany\\
 $^{3}$ Joint Institute for Nuclear Research, Dubna, 141980 Moscow
Region, Russia\\
 $^{4}$ Institute of Experimental Physics SAV, Watsonova 47, 
04353 Kosice, Slovakia}
 
\date{\today}
\begin{abstract}

Recent data from RHIC for high-$p_T$ hadrons in gold-gold collisions raised again
the long standing problem of quantitatively understanding the Cronin effect, i.e.
nuclear enhancement of high-$p_T$ hadrons due to multiple interactions in nuclear
matter. In nucleus-nucleus collisions this effect has to be reliably calculated as
baseline for a signal of new physics in high-$p_T$ hadron production. The only
possibility to test models is to compare with available data for $pA$ collisions,
however, all existing models for the Cronin effect rely on a fit to the data to be
explained.  We develop a phenomenological description based on the light-cone
QCD-dipole approach which allows to explain available data without fitting to them
and to provide predictions for $pA$ collisions at RHIC and LHC. We point out that
the mechanism causing Cronin effect drastically changes between the energies of
fixed target experiments and RHIC-LHC. High-$p_T$ hadrons are produced
incoherently on different nucleons at low energies, whereas the production
amplitudes interfere if the energy is sufficiently high. 

\smallskip
 
PACS: 24.85.+p, 13.85.Ni, 25.40.Qa 

\end{abstract}
\maketitle
 
It was first observed back in 1975 \cite{cronin1} that high-$p_T$ hadrons are not
suppressed in proton-nucleus collisions, but produced copiously.  This effect
named after James Cronin demonstrates that bound nucleons cooperate producing
high-$p_T$ particles. Indeed, it has been soon realized that multiple interactions
which have a steeper than linear $A$-dependence lead to the observed enhancement.
An adequate interpretation of the Cronin effect has become especially important
recently in connection with data from RHIC for high-$p_T$ hadron production in
heavy ion collisions \cite{phenix,star}. The observed suppression factor can be
understood as a product of two terms. One is due to multiple interactions within
the colliding nuclei, analogous to the Cronin effect. The second factor arises
from final state interaction with the produced medium, the properties of which are
thus probed. This second factor, the main goal of the experiment, can be extracted
from data only provided that the Cronin effect for nuclear collisions can be
reliably predicted. However, in spite of the qualitative understanding of the
underlying dynamics of this effect, no satisfactory quantitative explanation of
existing $pA$ data has been suggested so far. Available models contain parameters
fitted to the data to be explained (e.g. see \cite{ochiai,wang,levai}) and miss
important physics.  In this paper we suggest a comprehensive description of the
dynamics behind the Cronin effect resulting in parameter-free predictions which
agree with available data. 

First of all, the mechanism of multiple interactions significantly changes with
energy. At low energies a high-$k_T$ parton is produced off different nucleons
incoherently, while at high energies it becomes a coherent process. This is controlled
by the coherence length
 \beq
l_c = \frac{\sqrt{s}}{m_Nk_T}\ ,
\label{10}
 \eeq
 where $k_T$ is the transverse momentum of the parton produced at mid 
rapidity and then hadronizing into the detected hadron with transverse 
momentum $p_T$.

For a coherence length which is shorter than the typical internucleon
separation, the projectile interacts incoherently with individual
nucleons, just as for e.g. $pp$ scattering. However, QCD factorization is
violated by multiple scattering as discussed e.g. in \cite{bbl}.
Therefore, broadening of transverse momentum caused by initial/final
interactions, should not be translated into a modification of the parton
distribution of the nucleus if the coherence length is short.  In the
opposite limit, i.e. if the coherence length is longer than the nuclear
radius $R_A$, factorization applies. All amplitudes interfere coherently
and result in a collective parton distribution of the nucleus. This
difference is present in all of the various manners in which such
interactions are discussed. It is e.g. adequate to view a nucleus in the
nucleus momentum frame as a cloud of partons. Those with small $x$
overlap and are no longer associated with any individual nucleon. Small
$x$ corresponds to a long $l_c\sim 1/(xm_N)$. Again, factorization
applies, but the nuclear parton distribution is modified. The mean
transverse momentum of gluons increase \cite{mv} since their density
saturates at small $k_T$ \cite{glr,m90}.

 \medskip
 {\it Short coherence length}.  Broadening of transverse momentum of a
projectile parton propagating through a nuclear medium is quite a
complicated process involving rescatterings of the parton accompanied by
gluon radiation.  Apparently, this process involves soft interactions and
cannot be calculated perturbatively. Instead, one should rely on
phenomenology. Corresponding calculations have been performed in
\cite{jkt} in the framework of the light-cone QCD dipole approach. The
transverse momentum distribution of partons after propagation through
nuclear matter of thickness $T_A(b)=\int_{-\infty}^{\infty}
dz\,\rho_A(z)$ (the nuclear density integrated along the parton
trajectory at impact parameter $b$) has the form \cite{jkt},
 \beqn
\frac{dN_q}{d^2k_T} &=&
\int d^2r_1\,d^2 r_2\,e^{
i\,\vec k_T\,(\vec r_1 - \vec r_2)}\,
\Omega^q_{in}(\vec r_1,\vec r_2)\nonumber\\
&\times& e^{-{1\over2}\,
\sigma^N_{\bar qq}(\vec r_1-\vec r_2,x)\,T_A(b)}\ .
\label{20}
 \eeqn
 Here $\Omega^q_{in}(\vec r_1,\vec r_2)$ is the density matrix describing 
the impact parameter distribution of the quark in the incident hadron,
 \beq
\Omega^q_{in}(\vec r_1,\vec r_2) = 
\frac{\la k_0^2\ra}{\pi}\,
e^{-{1\over2}(r_1^2+r_2^2)
\la k_0^2\ra}\ ,
\label{30}
 \eeq
 where $\la k_0^2\ra$ is the mean value of the parton primordial
transverse momentum squared.

 The central ingredient of Eq.~(\ref{20}) is the phenomenological cross
section $\sigma^N_{\bar qq}(r_T,x)$ for the interaction of a nucleon with
a $\bar qq$ dipole of transverse separation $r_T$ at Bjorken $x$. In what
follows we use the simple parametrization \cite{gbw}
 \beq
\sigma_{\bar qq}(r_T,x)=\sigma_0\,\left[
1-e^{-{1\over4}\,r_T^2\,Q_s^2(x)}\right]\ ,
\label{40}
 \eeq
 the parameters of which were fixed by DIS data: $Q_s(x)=1\GeV \times
(x_0/x)^{\lambda/2}$ and $\sigma_0=23.03\mb$; $\lambda=0.288$;
$x_0=3.04\cdot 10^{-4}$.

Note that the $k_T$ distribution of quarks from a single $q-N$ scattering
process is not singular at $k_T\to 0$, but according to (\ref{40}) has a
Gaussian shape.  The phenomenon of saturation for soft gluons
\cite{glr,mueller} is the driving idea of parametrization \cite{gbw}.
Therefore, the mean momentum transfer in each scattering is not small,
but of the order of the saturation scale $Q_s(x)$.

Of course, for projectile gluons the broadening is stronger than for
quarks and the dipole cross section Eq.~(\ref{40}) should be replaced by
the glue-glue one $\sigma^N_{GG}={9\over4}\sigma^N_{\bar qq}$.

Besides broadening of transverse momentum, initial state interactions
also lead to energy loss \cite{eloss,kn}. While induced energy loss in
cold nuclear medium is negligibly small \cite{baier,eloss}, energy loss
due to hadronization in inelastic scattering reactions (which is
basically the same as for hadronization in vacuum) is important. The
first inelastic interaction of the incident hadron triggers energy loss
and the parton participating in the high-$p_T$ process arrives with a
noticeably reduced energy \cite{kn,eloss}.  We fixed the energy loss
$\Delta E$ to a mean value corresponding to the mean path length
calculated in \cite{eloss} and a rate of energy loss
$dE/dz=-2.5\GeV/\fm$.

For the cross section of $pA\to hX$ at high $p_T$ we use the standard
convolution expression based on QCD factorization \cite{3f},
 \beq
\sigma^{l_c\ll R_A}_{pA}(p_T) = \sum\limits_{i,j,k,l}
\widetilde F_{i/p} \otimes F_{j/A} 
\otimes \hat\sigma_{ij\to kl}\,
\otimes D_{h/k}\ ,
\label{60}
 \eeq
 where $F_{i/p}$ and $F_{j/A}$ are the distributions of parton species
$i,j$ in Bjorken $x_{1,2}$ and transverse momentum in the colliding
proton and nucleus respectively. However, to describe the nonfactorizable
multiple interactions the beam parton distribution $\widetilde F^p_i$ is
modified by by the transverse momentum broadening Eq.~(\ref{20}) and by
shifting $x_1$ to $\tilde x_1 = x_1 + \Delta E/(x_1 E_p)$. For $\la
k_0^2\ra$ in (\ref{30}) we use the next-to-leading value from \cite{wang}
fitted to data for hadron production in $pp$ collisions.  For the parton
distribution functions in a nucleon we use the leading order GRV
parametrization \cite{grv}.  The nuclear parton distribution, $F_{j/A}$,
is unchanged compared to a free nucleon, except at large $x_2$ where it
is subject to medium modifications (EMC effect) which are parametrized
according to \cite{eks}. For the hard parton scattering cross section
\cite{3f} we use regularization masses $m_G=0.8\GeV$ and $m_q=0.2\GeV$
for gluon and quark propagators respectively. Such a large effective
gluon mass was introduced to reproduce the strong nonperturbative
light-cone gluon interaction \cite{kst2} dictated by diffraction data.
The fragmentation functions of a parton $k$ into the final hadron $h$,
$D_{h/k}$ are taken from \cite{D} in leading order. We use the realistic
Woods-Saxon parametrization for the nuclear density.

\begin{figure}[tbh]
\centerline{\psfig{figure=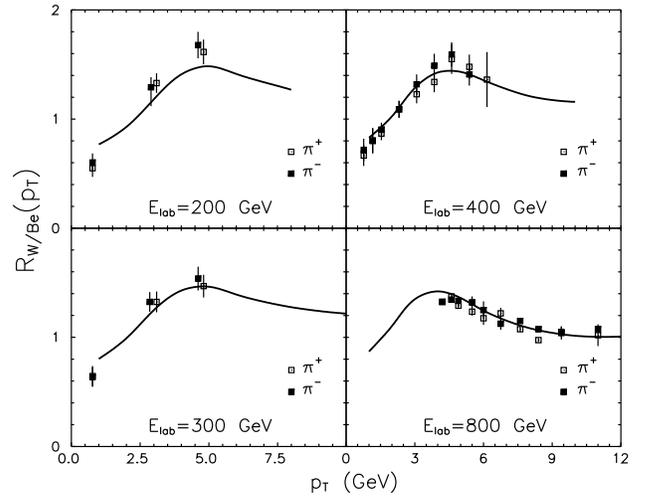,width=8cm}}
  \protect\caption{Ratio of the charged pion production cross sections
for tungsten and beryllium function of the transverse momentum of the
produced pions. The curves correspond to the parameter-free calculation
Eq.~(\ref{60}), the data are from fixed target experiments
\cite{cronin2,e605}}
 \label{fixed-target} 
 \end{figure}

As far as all the parameters in (\ref{60}) are fitted to data for proton
target, we have no further adjustable parameters and can predict nuclear
effect. The results of parameter-free calculations for the production of
charged pions are compared in Fig.~\ref{fixed-target} with fixed target
data. $R_{W/Be}(p_T)$ is the ratio of the tungsten and beryllium cross
sections at $200-400\GeV$ \cite{cronin2} and $800\GeV$ \cite{e605} as
function of $p_T$.

\medskip {\it Long coherence length}. In the limit of $l_c\gg R_A$ a hard
fluctuation in the incident proton containing a high-$p_T$ parton
propagates through the whole nucleus and may be freed by the interaction.
Since multiple interactions in the nucleus supply a larger momentum
transfer than a nucleon target, they are able to resolve harder
fluctuations, i.e. the average transverse momentum of produced hadrons
increases. In this case broadening looks like color filtering rather than
Brownian motion.

Instead of QCD factorization we employ the light-cone dipole formalism in
the rest frame of the target which leads to another factorized
expression, valid at $x_2\ll1$,
 \beq
\sigma^{l_c\gg R_A}_{pA}(p_T) = F_{G/p}\otimes
\sigma(GA\to G_1G_2X)\otimes D_{h/G_1}\ .
\label{70}
 \eeq 
 We assume that high-$p_T$ hadrons originate mainly from radiated gluons
at such high energies. The cross section of gluon radiation reads
\cite{km,kst1,kst2},
 \beqn
&& \frac{d\sigma(GA\to G_1G_2X)}
{d^2p_T\,dy_{1}} =
\int d^2b\int d^2r_1d^2r_2\,
e^{i\vec p_T(\vec r_1-\vec r_2)}\,
\nonumber\\ &\times& 
\overline{\Psi_{GG}^*(\vec r_1,\alpha)
\Psi_{GG}(\vec r_2,\alpha)}
\left[1 - e^{-{1\over2}\sigma^N_{3G}(r_1,x)T_A(b)}
\right.\nonumber\\ &-& \left.
e^{-{1\over2}\sigma^N_{3G}(r_2,x)T_A(b)} +
e^{-{1\over2}\sigma^N_{3G}(\vec r_1-\vec r_2,x)T_A(b)} 
\right]\ .
\label{80}
 \eeqn
 Here $\alpha = p_+(G_1)/p_+(G)$ is the momentum fraction of the radiated
gluon; $\sigma^N_{3G}(r,\alpha)$ is the dipole cross section for a
three-gluon colorless system, where $\vec r$ is the transverse separation
of the final gluons $G_1$ and $G_2$. It can be expressed in terms of the
usual $\bar qq$ dipole cross sections,
 \beq
\sigma^N_{3G}(r) = {2\over9}\Bigl\{
\sigma_{\bar qq}(r) + \sigma_{\bar qq}(\alpha r) 
+ \sigma_{\bar qq}[(1-\alpha)r]\Bigr\}\ .
\label{90}
 \eeq

The light-cone wave function of the $G_1-G_2$ Fock component of the
incoming gluon including the nonperturbative interaction of the gluons
reads \cite{kst2},
 \beqn
&& \Psi_{GG}(\vec r,\alpha) = \frac{\sqrt{8\alpha_s}}{\pi\,r^2}\,
\exp\left[-\frac{r^2}{2\,r_0^2}\right]\,
\Bigl[\alpha(\vec e_1^{\,*}\cdot\vec e)(\vec e_2^{\,*}\cdot\vec r) 
\nonumber\\ &+&
(1-\alpha)(\vec e_2^{\,*}\cdot\vec e)(\vec e_1^{\,*}\cdot\vec r)-
\alpha(1-\alpha)(\vec e_1^{\,*}\cdot\vec e_2^{\,*})(\vec e\cdot\vec r)
\Bigr]\nonumber\ ,\\
\label{100}
 \eeqn
 where $r_0=0.3\fm$ is the parameter characterizing the strength of the
nonperturbative interaction which was fitted to data on diffractive $pp$
scattering. The product of the wave functions is averaged in (\ref{80})
over the initial gluon polarization, $\vec e$, and summed over the final
ones, $\vec e_{1,2}$. 

\begin{figure}[tbh] 
\includegraphics{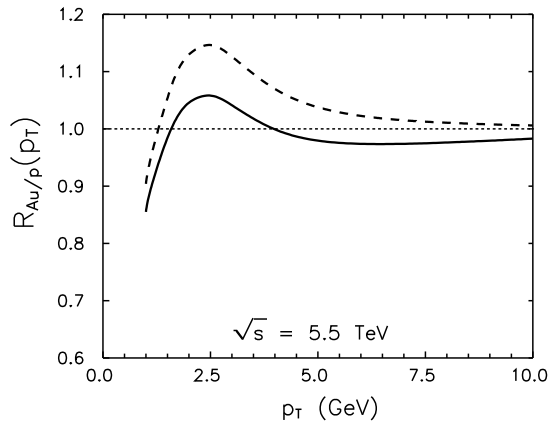} 
\begin{center} \vspace{5.5cm}
\parbox{8cm}
 {\caption[Delta]
 {Ratio of $p-Au$ to $pp$ cross sections as function 
of transverse momentum of produced pions at the energy of LHC calculated 
with Eq.~(\ref{80}).
The dashed and solid curves correspond to calculations 
without and with gluon shadowing respectively.} 
\label{lhc}} 
\end{center} 
 \end{figure}

Expression (\ref{80}) with the exponentials expanded to first order in
the nuclear thickness also provides the cross section for gluon radiation
in $pp$ collisions. This cross section reproduces well the measured pion
spectra in $pp$ collisions. The results for the ratio of pion production
rates in $pA$ and $pp$ collisions obtained using
Eqs.~(\ref{70})-(\ref{80}) for mid rapidity at the energy of LHC,
$\sqrt{s}=5.5\TeV$ are shown by curve in Fig.~\ref{lhc}.

Note that at the high LHC energy the eikonal formula Eq.~(\ref{80})  is
not exact. The higher Fock components $|3G\ra,\ |4G\ra$, etc. lead to
additional corrections called gluon shadowing. These fluctuations are
heavier than $|2G\ra$, correspondingly, the coherence length is shorter,
and one should sum over all different trajectories of the gluons.  This
problem was solved in \cite{kst2,knst,krtj} and a suppression factor
$R_G(x,Q^2,b)$ due to gluon shadowing was derived. Here we make use of
those results replacing the dipole cross sections in (\ref{80}),
$\sigma_{3G}$ by $R_G\,\sigma_{3G}$. This suppression factor leads to a
reduction of the Cronin effect as is demonstrated by the solid curve in
Fig.~\ref{lhc}. Note that this curve approaches unity from below at high
$p_T$.

\medskip {\it Predictions for RHIC}. The calculations in the energy range
of RHIC are most complicated since this is the transition region between
the regimes of long (small $p_T$) and short (large $p_T$) coherence
lengths. One can deal with this situation relying on the light-cone Green
function formalism \cite{krt1,krt2,knst}. However, in this case the
integrations involved become too complicated. Fortunately, the coherence
length at the energy of RHIC is rather long, $l_c\sim 5\fm$, within the
$p_T$-range where the Cronin effect has an appreciable magnitude.  
Therefore, the corrections to the asymptotic expression Eq.~(\ref{70})
should not be large and can be approximated by linear interpolation
performed by means of the the so called nuclear longitudinal formfactor
$F_A(q_c,b)$ \cite{kp,eloss},
 \beqn 
\sigma_{pA}(p_T) &=& 
\int d^2b\,\left\{\Bigl[1-\la F^2_A(q_c,b)\ra\Bigr]\, 
\sigma^{l_c\ll R_A}_{pA}(p_T,b)\right.
\nonumber\\ &+& \left. 
\la F^2_A(q_c,b)\ra\, 
\sigma^{l_c\gg R_A}_{pA}(p_T,b)\right\}\ . 
\label{110}
 \eeqn
 Here $\sigma_{pA}(p_T,b)$ is the unintegrated $\vec b$-dependent
contribution to the cross section $\sigma_{pA}(p_T)$,
 \beq
F_A(q_c,b)= \frac{1}{T_A(b)}
\int\limits_{-\infty}^{\infty}
dz\,\rho_A(b,z)\,e^{iq_cz}\ ,
\label{120}
 \eeq
 where $q_c=1/l_c$. The formfactor is averaged weighted with the cross 
section at fixed $p_T$ and varying initial and final parton momenta.

\begin{figure}[tbh] 
\includegraphics{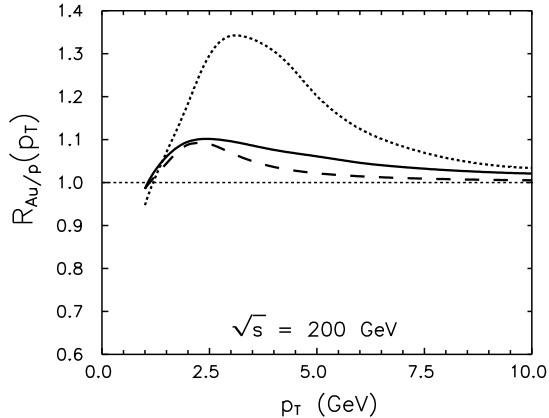} 
\begin{center} \vspace{5.5cm}
\parbox{8cm}
 {\caption[Delta]
 {Predictions for RHIC. The dotted and dashed curves are calculated 
at $\sqrt{s}=200\GeV$ using Eqs.~(\ref{60}) and (\ref{70}) respectively. 
The final prediction
taking into account the coherence length is shown by the solid curve.} 
\label{rhic}} 
\end{center} 
 \end{figure}

Expression (\ref{110}) interpolates between the cross sections
$\sigma^{l_c\ll R_A}_{pA}(p_T)$, Eq.~(\ref{60}), and $\sigma^{l_c\gg
R_A}_{pA}(p_T)$, Eq.~(\ref{70}), which are shown in Fig.~\ref{rhic} by
dotted and dashed curves respectively.
 It is interesting that the dashed curve exposes a weaker nuclear 
enhancement than the dotted one. This might be interpreted as 
Landau-Pomeranchuk suppression of the radiation spectrum compared to the 
Bethe-Heitler regime.

 Our prediction for $\sqrt{s}=200\GeV$ calculated with Eq.~(\ref{110}) is
depicted by the solid curve which nearly coincides with the $l_c\ll R_A$
one at $p_T < 2\GeV$ and is rather close to it at higher $p_T$.  
$\l_c\gg R_A$ regime at higher $p_T$. Eventually, all three curves
approach 1 at large $p_T>10\GeV$.

No sizeable gluon shadowing is expected at RHIC energy. The reason is
that the effective coherence length for gluon shadowing evaluated in
\cite{krt2} is nearly an order of magnitude shorter than $l_c$ for single
gluon radiation as given by (\ref{10}).

\medskip {\it Summary:} the mechanism of high-$p_T$ hadron production has two
limiting regimes.  At $l_c\ll R_A$ a high-$p_T$ particle is produced incoherently
on different nucleons, and the Cronin effect is due to soft multiple initial/final
state interactions which break QCD factorization.  On the contrary, for $l_c\gg
R_A$ the process of gluon radiation takes long time even for high transverse
momenta. As a result, coherent radiation from different nucleons is subject to
Landau-Pomeranchuk suppression. Using the light-cone dipole approach we provided
the first parameter-free calculations for the Cronin effect in $pA$ collisions,
i.e. no fit is done to the data to be described. Our results agree well with
available data and we provided predictions for high-$p_T$ pion production at RHIC
and LHC. 

\bigskip

\noindent
 {\bf Acknowledgment}: we are grateful to J\"org H\"ufner, Mikkel Johnson
and J\"org Raufeisen for stimulating discussions. This work has been
partially supported by a grant from the Gesellschaft f\"ur
Schwerionenforschung Darmstadt (GSI), grant No.~GSI-OR-SCH. The work of
J.N. has been supported in part by the Slovak Funding Agency, Grant No.
2/1169 and Grant No. 6114.

\end{document}